\documentstyle[twoside,fleqn,espcrc2,psfig]{article}

\newcommand{\AmS}{{\protect\the\textfont2
  A\kern-.1667em\lower.5ex\hbox{M}\kern-.125emS}}

\title{Relationship between X-ray and ultraviolet emission in 3C 273}

\author{S.\ Paltani\address{Geneva Observatory\\
                            51, ch.\ des Maillettes\\
                            CH-1290 Sauverny, Switzerland}%
\address{INTEGRAL Science Data Centre\\
                            16, ch.\ d'Ecogia\\
                            CH-1290 Versoix, Switzerland},%
        T.J.-L.\ Courvoisier$^{\mathrm{ab}}$,
        M.\ T\"urler$^{\mathrm{ab}}$,
    and R.\ Walter$^{\mathrm{b}}$}
       
\begin{document}

\begin{abstract}
  In 3C 273, ultraviolet flux and X-ray flux measured by BATSE are not
  well correlated, contrarily to predictions of several models, unless
  the X-ray flux lags the UV emission by 1.75 yr. The absence of
  observed correlation at small lag cannot be due to spectral
  variability. A Comptonizing corona model is however compatible with
  all UV and X-ray observations covering the BATSE period.
\end{abstract}

\maketitle

\section{INTRODUCTION}
Many models predict some physical links between the X-ray emission and
the ultraviolet emission in AGN. It is however very difficult to test
these predictions, mostly because it requires an intense X-ray
monitoring. We make a first step in that direction by comparing the
X-ray emission of 3C 273 observed by BATSE with the ultraviolet
observations performed by the IUE satellite. We explore here whether
BATSE data, which are quasi-continuous in time and cover a large
spectral domain, can provide useful constraints. We use simple
phenomenological models that predict - at least partly - how X-ray and
UV emissions interplay, and test whether these models can account for
the observed data.

\section{OBSERVATIONS}
\begin{figure}
\mbox{\psfig{bbllx=16pt,bblly=100pt,bburx=600pt,bbury=764pt,file=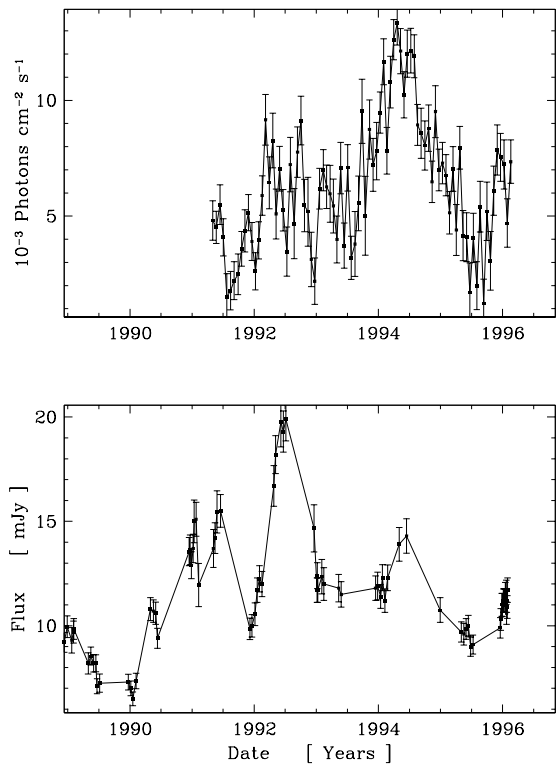,width=8.1cm}}
\caption{Top: BATSE light curve binned into 20-day periods. Bottom: Part of the UV
light curve from the IUE satellite. The points are connected only to enhance the structures.}
\end{figure}
We consider here the continuous BATSE observations of 3C 273. A photon
flux between 20 keV and 200 keV is available from COSSC, which is
obtained by the fit of a power-law with an index of 1.7 (Fig.\ 1
top). The most obvious feature is a large bump between 1994 and 1995,
but other peaks exist, in particular in 1992-1993. We shall compare
this light curve with the UV light curve obtained with the IUE
satellite between 1250 and 1300 Å (Fig.\ 1 bottom). Among others, two
bumps are clearly present in 1992-1993 and in 1994-1995, in periods
roughly corresponding to the bumps in the BATSE light curve; however,
the relative intensities of the bumps are completely different: the
peak between 1992 and 1993 is strong with IUE, and weak with BATSE;
while the peak between 1994 and 1995 is weak with IUE, but strong with
BATSE. If the two components are physically related without delay, and
if the BATSE flux is a good measure of the total X-ray flux, it
implies that a third parameter must play a role in the physical
relationship.

\section{BATSE-IUE CROSS-CORRELATION}
\begin{figure}
\mbox{\psfig{bbllx=36pt,bblly=220pt,bburx=775pt,bbury=721pt,file=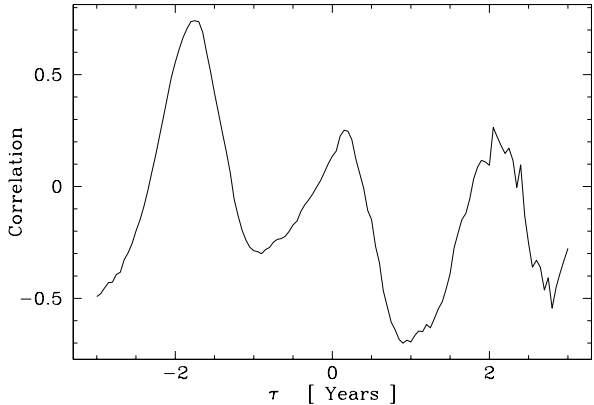,width=7.5cm}}
\caption{Correlation between the IUE UV light curve and the BATSE X-ray light curve.}
\end{figure}
We calculate the cross-correlation of the UV and X-ray light curves
(Fig.\ 2) by interpolating the BATSE light curve, to quantify the
relationship between the light curves. There is a very significant
peak at a delay around -1.75 yr. This would mean that a UV-flux
increase is followed by a X-ray flux increase a bit less than 2 years
later. The correlation is however dominated by the presence of a
single dominating bump in both light curves, and is possibly only a
coincidence. However, if the two bumps are signatures of a unique
event, either the distance between the UV and X-ray emission zones
should be larger than one light-year, or the interaction travels at a
speed much smaller than the speed of light.

The correlation has a local maximum close to $\tau=0$, but at a value
that makes it hardly significant. If there exists any physical
relationship between the UV and X-ray emissions at small lags, other
parameters must appear in this relationship and contribute in a large
part to the ratio of UV and X-ray fluxes. In that case, the (strongly
damped) correlation peak suggests that the UV light curve follows the
X-ray light curve with a delay of ~2 months.  Such a delay would be
natural for a reprocessing model, where the accretion disk is heated
by the X-ray radiation from a source located above the accretion disk.


\section{THE CORRELATION AT -1.75 yr}
\begin{figure}
\mbox{\psfig{bbllx=0pt,bblly=50pt,bburx=533pt,bbury=232pt,file=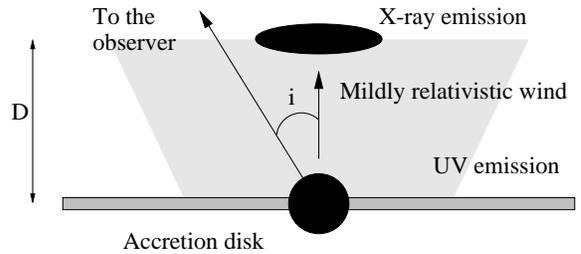,width=7.5cm}}
\caption{Sketch of the wind and shock model: the power of a mildly relativistic wind
  is transformed into X-ray emission.}
\end{figure}
If the large correlation peak around -1.75 yr is real, it is best
explained by models where the X-ray source is located far from the UV
source.  Courvoisier \& Camenzind \cite{CoCa89} have proposed a
scenario (illustrated on Fig.\ 3) that can account easily for this lag.

A mildly-relativistic wind is shocked at a distance ~1 pc above the
disk. We assume that the power of the wind is correlated to the
blue-bump luminosity. The power of the wind reaches the shock region
after a delay $D/v$, where $D \simeq 1$ pc is the distance between the
accretion disk and the shock zone, and $v$ is the velocity of the wind.
The observed delay between the lightcurves is then:
\begin{equation}
\tau=(1+z)\cdot\left[ \frac{D}{v}-\frac{D}{c}\cdot(1-\frac{i^2}{2}) \right]
\end{equation}
where $i \simeq 0.25$ rad is the angle between the observer and the
perpendicular to the accretion disk (deduced from superluminal motion)
and gives a negligible contribution, $c$ is the speed of light,
$z=0.158$ is the redshift. The observed delay is obtained for $v
\simeq 2/3 c$, compatible with the assumption of mildly relativistic
jet, although several times larger that the value used in
\cite{CoCa89}. It must be noted that the interaction cannot travel at
the speed of light.

\section{CORRELATION WITHOUT DELAY}
\begin{figure}
\mbox{\psfig{bbllx=16pt,bblly=220pt,bburx=600pt,bbury=721pt,file=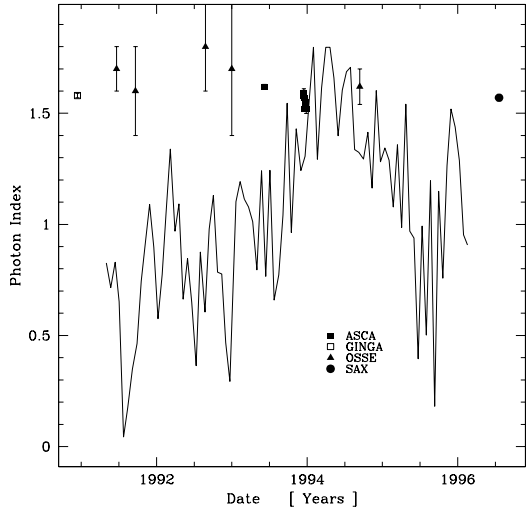,width=7.5cm}}
\caption{Best fit photon index curve in the case of correlated UV and X-ray energy fluxes compared with historic measurements of the photon index.}
\end{figure}
An obvious parameter that can affect a correlation between UV flux and
the X-ray flux derived from BATSE is spectral variability. Several
measurements of the spectral index of 3C 273 can be found in the
litterature. The best measurement up to now has been made by SAX
\cite{Gr..97}, which found that the X-ray spectrum of 3C 273 from 1 to
200 keV follows a power-law with an index 1.57$\pm$ 0.01. OSSE
measurements \cite{Jo..95,Mc..97} do not reach this level of accuracy,
but low-energy experiments, like EXOSAT, GINGA, and ASCA, have shown
that a small spectral variability exists \cite{MaBa94,CaMa97}. Taking
this spectral variability into account, it means that the BATSE flux
is not necessarily representative of the total X-ray flux. A
distinction must be made however between the energy flux, and the
photon flux. For a given 20--200 keV flux, the effect of a hardening
of the X-ray spectrum may be an increase of the total energy flux, but
a decrease of the total photon flux, as it is dominated by low-energy photons.

\begin{figure}
\mbox{\psfig{bbllx=16pt,bblly=220pt,bburx=600pt,bbury=721pt,file=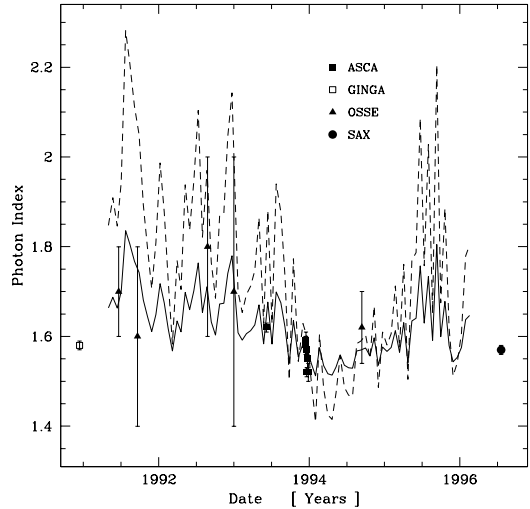,width=7.5cm}}
\caption{Best fit photon index curves in the case of correlated UV and X-ray photon
fluxes (dashed line) and in the case of the Comptonizing corona model (solid line) compared with historic measurements of the photon index.}
\end{figure}
If we assume that either the total X-ray energy flux or the total
X-ray photon flux is proportional to the UV flux, we can calculate how
the photon index would have to change to account for the observed
BATSE and UV fluxes. We take a X-ray power-law that extends from 1 keV
up to 1 MeV (other limits do not affect the qualitative results). The
calculated photon index curves can be fitted to the photon index
measurements from other missions. Fig.\ 4 shows that the assumption of
correlated X-ray and UV energy fluxes cannot give a good fit to the
observations.  Fig.\ 5 (dashed curve) shows that the assumption that
their photon fluxes are correlated requires very large, photon index
variations, which is not supported by the observations.  It is
therefore very improbable that the UV flux and the total X-ray flux
(either energy or photon) are completely correlated without (or with a
small) delay.

\section{COMPTONIZING CORONA}
\begin{figure}
\mbox{\psfig{bbllx=0pt,bblly=50pt,bburx=458pt,bbury=204pt,file=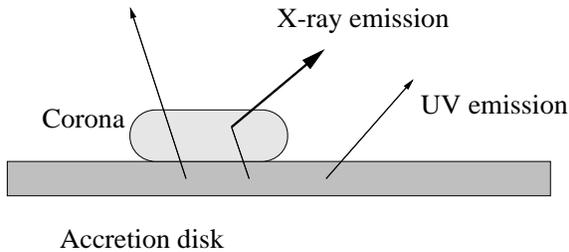,width=7.5cm}}
\caption{Comptonizing corona model: UV photons that interact with the corona
escape in the form of X-ray radiation}
\end{figure}
Walter \& Courvoisier \cite{WaCo90,WaCo92} have developed a model to
explain the peculiar relationship observed between the photon index
and the X-ray-to-UV photon flux ratio in several AGN, among which 3C
273. We apply here this model in a corona geometry, where the corona
covers only a part of the accretion disk, as shown in Fig.\ 6. A
thermal plasma of electrons above the disk Comptonizes the UV flux
from the accretion disk. The larger the optical depth of the plasma
and its covering factor, the larger the ratio of X-ray-to-UV photon
flux. If the corona has a small optical depth $\tau$, the spectral
index is related to $\tau$ and to the electron temperature
\cite{WaCo90,WaCo92}.

We can therefore deduce a relationship between the photon index and
the X-ray-to-UV photon flux ratio. We can also calculate in this case
the photon index curve predicted by this model, taking into account
the UV and BATSE light curves. We assume here that the mean photon
index is 1.6 and that the plasma temperature is $1.7\cdot 511$ keV, as
in \cite{WaCo92}. We use also the fact that SAX observed an unique
power-law from 1 to 200 keV.

The photon index curve implied by this model is shown on Fig.\ 5
(solid curve). It is similar to the one obtained in the case of
complete correlation, but with a much smaller dispersion. It fits
relatively well the historic measurements of the photon index by other
telescopes. Thus this model is compatible with all the UV and X-ray
observations covering the BATSE period.

\section{CONCLUSION}
Correlation analysis points to a X-ray source located far from the
ultraviolet source. This correlation however awaits confirmation, as
it can easily be due to a coincidence. If the correlation is physical,
it is most probable that interaction between the UV and X-ray emissions
travels at a speed slower than the speed of light. Wind and shock
model is in agreement with the observed delay, provided that the UV
light curve is indicative of the power transported by the wind.

Assuming a relationship without delay between X-ray and UV emission,
we find that models that predict a correlation between the X-ray
energy flux and the UV flux are hard to reconcile with the
observations presented here. In particular, reprocessing models fall
in this category. The same difficulty arises for models that predicts
a correlation between the UV and X-ray photon fluxes.

The Comptonizing corona model discussed here predicts a relationship
between the X-ray photon flux, the UV photon flux, and the X-ray
photon index. The set of observations discussed here actually supports
this model.

\end{document}